\documentclass[%
 reprint,
superscriptaddress,
 amsmath,amssymb,
 aps,
 prl,
]{revtex4-2}
\usepackage{graphicx}
\usepackage{dcolumn}
\usepackage{bm}
\usepackage{float}
\usepackage{algorithm}
\usepackage{algpseudocode}
 \usepackage{relsize}
 
\usepackage{lineno}  

\usepackage[breaklinks=true,colorlinks=true,linkcolor=blue,urlcolor=blue,citecolor=blue]{hyperref}
\usepackage[utf8]{inputenc}
\usepackage[T1]{fontenc}
\usepackage{mathptmx}
\usepackage{etoolbox}
\usepackage{comment}

\usepackage{color,soul}
\usepackage{xcolor}
\usepackage{physics} 
\usepackage{soul}
\usepackage{lipsum}

\begin{document}

\preprint{APS/123-QED}


\title{Capturing dynamical correlations using implicit neural representations}


%

\author{Sathya Chitturi}
 \email{chitturi@stanford.edu}
\affiliation{SLAC National Accelerator Laboratory, Menlo Park, CA 94025, USA}
\affiliation{Department of Materials Science and Engineering, Stanford University, Stanford, CA 94305, USA}

\author{Zhurun Ji}%
\email{zhurun@stanford.edu}
\affiliation{Department of Physics and Applied Physics, Stanford University, Stanford, CA 94305, USA}
\affiliation{Geballe Laboratory for Advanced Materials, Stanford University, Stanford, CA 94305, USA}

\author{Alexander\,N. Petsch}
\email{apetsch@stanford.edu}
\affiliation{SLAC National Accelerator Laboratory, Menlo Park, CA 94025, USA}
\affiliation{Stanford Institute for Materials and Energy Sciences, Stanford University, Stanford, CA 94305, USA}
\affiliation{H.H. Wills Physics Laboratory, University of Bristol, Bristol BS8 1TL, United Kingdom}

\author{Cheng Peng}%
 \affiliation{Stanford Institute for Materials and Energy Sciences, Stanford University, Stanford, CA 94305, USA}

\author{Zhantao Chen}%
 \affiliation{Stanford Institute for Materials and Energy Sciences, Stanford University, Stanford, CA 94305, USA}

\author{Rajan Plumley}
\affiliation{SLAC National Accelerator Laboratory, Menlo Park, CA 94025, USA}
\affiliation{Stanford Institute for Materials and Energy Sciences, Stanford University, Stanford, CA 94305, USA}
\affiliation{Department of Physics, Carnegie Mellon University, Pittsburgh, PA 15213, USA}

\author{Mike Dunne}
\affiliation{SLAC National Accelerator Laboratory, Menlo Park, CA 94025, USA}

\author{S.\ Mardanya}
\affiliation{Department of Physics and Astrophysics, Howard University, Washington, USA}

\author{S.\ Chowdhury}
\affiliation{Department of Physics and Astrophysics, Howard University, Washington, USA}

\author{H.\ Chen}
\affiliation{Department of Physics, Northeastern University, Boston, USA}

\author{A.\ Bansil}
\affiliation{Department of Physics, Northeastern University, Boston, USA}

\author{A.\ Feiguin}
\affiliation{Department of Physics, Northeastern University, Boston, USA}

\author{A.\,I. Kolesnikov}
\affiliation{Neutron Scattering Division, Oak Ridge National Laboratory, Oak Ridge, Tennessee 37831, USA}

\author{D. Prabhakaran}
\affiliation{Department of Physics, University of Oxford, Clarendon Laboratory, Oxford OX1 3PU, United Kingdom}

\author{S.\,M. Hayden}
\affiliation{H.H. Wills Physics Laboratory, University of Bristol, Bristol BS8 1TL, United Kingdom}

 \author{Daniel Ratner}
\affiliation{SLAC National Accelerator Laboratory, Menlo Park, CA 94025, USA}

 \author{Chunjing Jia}%
 \affiliation{SLAC National Accelerator Laboratory, Menlo Park, CA 94025, USA}
\affiliation{Stanford Institute for Materials and Energy Sciences, Stanford University, Stanford, CA 94305, USA}
\affiliation{Department of Physics, University of Florida, Gainesville, FL 32611, USA}

 \author{Youssef Nashed}
 \affiliation{SLAC National Accelerator Laboratory, Menlo Park, CA 94025, USA}

 \author{Joshua J. Turner}
 \email{joshuat@slac.stanford.edu}
\affiliation{SLAC National Accelerator Laboratory, Menlo Park, CA 94025, USA}
\affiliation{Stanford Institute for Materials and Energy Sciences, Stanford University, Stanford, CA 94305, USA}

\date{\today}

\begin{abstract}
The observation and description of collective excitations in solids is a fundamental issue when seeking to understand the physics of a many-body system. Analysis of these excitations is usually carried out by measuring the dynamical structure factor, $S(\mathbf{Q}, \omega)$, with inelastic neutron or x-ray scattering techniques and comparing this against a calculated dynamical model. Here, we develop an artificial intelligence framework which combines a neural network trained to mimic simulated data from a model Hamiltonian with automatic differentiation to recover unknown parameters from experimental data. We benchmark this approach on a Linear Spin Wave Theory (LSWT) simulator and advanced inelastic neutron scattering data from the square-lattice spin-1 antiferromagnet La$_2$NiO$_4$. We find that the model predicts the unknown parameters with excellent agreement relative to analytical fitting. In doing so, we illustrate the ability to build and train a differentiable model only once, which then can be applied in real-time to multi-dimensional scattering data, without the need for human-guided peak finding and fitting algorithms. This prototypical approach promises a new technology for this field to automatically detect and refine more advanced models for ordered quantum systems.
\end{abstract}

\maketitle


\section{\label{sec:level1}Introduction}
Quantum matter, as featured by the existence of macroscopic order from microscopic spin or charge arrangements or phases with spontaneous symmetry breaking, represents an abundant and complex class of materials in condensed matter physics. For example, the magnetic configuration of a material and its dynamics are often a synergistic effect of multiple interactions as well as crystalline symmetries. The collective spin excitations in most magnetic materials, such as spin waves or magnons, act as probes of those interactions. This information is reflected in their dispersion relations and correlations, and possesses a wide range of potential applications, which include forming spintronics devices, as well as carrying, transferring, and storing information~\cite{chumak2015magnon,neusser2009magnonics,gutfleisch2011magnetic}.

In the last few decades, a big aim has been to characterize these excitations, and this has been facilitated by advances in spectroscopic techniques, such as neutron scattering techniques~\cite{rossat1991neutron,chatterji2005neutron,braden2002inelastic,coldea2001spin}. These techniques use the energy and directions of scattered neutrons to measure the dispersion relations, lifetimes, and amplitudes of the spin excitations. While neutron scattering could provide valuable information about the structure and magnetic properties of materials, the limited availability of neutrons, i.e. low flux compared to other scattering techniques, the small partial differential cross section, and the complexity of interpretation, has made the question how to effectively enhance the efficiency of these experiments a long term topic~\cite{weinfurther2018model,peterson2018advances}. Moreover, the interpretation of neutron scattering data can be challenging and time-consuming due to the complex nature of this physical process, the diversity of samples, and limited knowledge from theoretical modeling. With these obstacles, there is a need for deep and synergistic collaboration among experiment, theory, and data analysis to accelerate and simplify the understanding of spin properties~\cite{chen2021machine}. 

Moreover, as rates of data collection increase as well as the ability to collect hyper-dimensional datasets, it is important to be able to readily allow for real-time fitting during experimentation. The ability to perform `on-the-fly' fitting~\cite{li2015molecular} can allow for efficient use of expensive beamtime by knowing when sufficient data is collected, as well as by coupling to adaptive sampling methods to gain the most information about parameters of interest with the least number of measurements possible. Currently, for neutron scattering data, real-time fitting can require substantial preparation -- for example, direct fitting with the package known as SpinW~\cite{toth2015linear} requires the extraction of the eigenmodes of the system and therefore, needs an accurate and favorably automatic peak extraction algorithm. When the chosen paths in reciprocal space are numerous or the dispersion relations change significantly along those paths, this can involve significant human guidance and monitoring. In addition, fitting directly through SpinW does not take into consideration the magnon peak intensities or their shapes. Approaches to fit peak intensities and shapes directly, such as Multi- or Tobyfit implemented in HORACE~\cite{ewings2016horace}, are possible alternatives -- however, these fitting procedures still require significant human guidance and are either extremely slow and therefore incompatible with data acquisition rates or else they require an analytical, rapidly calculable spin wave model. Finding such a model is usually only feasible for simpler systems with minimal frustration or a low number of magnetically distinct sites.

In this work, we solve this problem by developing a machine learning platform based on a data-driven differentiable neural representation (interchangeably referred-to as the "surrogate model") which will dramatically affect how these types of experiments are carried out. This method works with any algorithm, magnetic or non-magnetic, that calculates the dynamical structure factor $S(\mathbf{Q}, \omega)$, where $\mathbf{Q}$ is the scattering vector and $\hslash\omega$ is the energy transfer. The platform is based on recent developments in implicit neural representation theory \cite{sitzmann2020implicit,xie2022neural}, and the dynamical structure factor is the general function measured in many inelastic x-ray and neutron experiments and is related to the partial differential cross section by $\tfrac{\mathrm{d}^2\sigma}{\mathrm{d}\Omega\mathrm{d}E_f}=k_f/k_i\ S(\mathbf{Q}, \omega)$, where $k_i$ and $k_f$ are the incident and final wave vector. 
Here, the dynamical structure factor is approximated to $S(\mathbf{Q},\omega)\propto \sum_{m,n}\int \mathrm{d}t\ e^{-i\mathbf{Q}\cdot(\mathbf{r}_{m}-\mathbf{r}_{n})}e^{-i\omega t}\langle S_{m}(t)S_{n}(0)\rangle$, where $\langle S_{m}(t)S_{n}(0)\rangle$ represents spin-spin correlations at different atomic sites $m,n$. The neutron polarization factor as well as the magnetic form factor are neglected here. This work extends recent prior work which used machine learning methods to calculate the static structure factor after training on simulated training data $S(\mathbf{Q})$~\cite{samarakoon2020machine,samarakoon2022integration}. 

Our focus in this work is the introduction of neural implicit representations to model the dynamical regime. Convention feed-forward neural networks, trained on large linear spin wave theory (LSWT) simulations, have also been applied to inelastic neutron scattering data to predict the type of magnetic exchange from data~\cite{butler2021interpretable}. More recently, a cycle-GAN approach, which is able to make experimental data look like simulated data has been applied as a pre-processing step to further aid in predicting a certain class of spin-model~\cite{anker2022using}. We note that our work is complementary to these approaches in that our focus is on predicting continuous parameters within an assumed Hamiltonian model rather than predicting the discrete choice of functional form. For a comprehensive survey of the application of machine learning methods to x-ray and neutron data, please refer to Ref.~\cite{chen2021machine}. 

Our approach offers a straightforward method to both predict and quickly learn parameters of a spin or charge Hamiltonian directly from inelastic neutron or resonant X-ray scattering data. The power of this approach will be especially impactful in using expensive and advanced computational methods for simulating strongly coupled electrons, such as exact diagonalization (ED)~\cite{dagotto1994correlated}, density matrix renormalization group (DMRG)~\cite{White1992, white1993density}, determinant quantum Monte-Carlo (DQMC)~\cite{dqmc1,dqmc2}, and variational Monte Carlo (VMC)~\cite{ferrari2018dynamical, hendry2021chebyshev}. 

To further demonstrate the versatility of our method, we report the results using a less expensive series of calculations based on mean field theory, the linear spin wave theory (LSWT) framework~\cite{kubo1952spin}. We simulate massive LSWT spectra from a spin-1 square-lattice Heisenberg Hamiltonian model for a large phase space of Hamiltonian parameters. We use a GPU-based machine learning framework to learn how to recover these parameters from scattering data and, in particular, show the method does not rely on peak fitting algorithms for experimental data. We test our approach on experimental time-of-flight neutron spectroscopy data \cite{petsch2023highenergy} taken on the quasi-2D N\'eel antiferromagnet La$_2$NiO$_4$ at the Spallation Neutron Source at the Oak Ridge National Laboratory~\cite{granroth2010sequoia} and show that the approach can utilize raw data almost directly from neutron scattering files containing -- $\mathbf{Q}, \omega$, and $S(\mathbf{Q}, \omega)$ -- to return Hamiltonian parameters that represent the system under study. In addition, we use a Monte-Carlo simulation of the experimental data collection to demonstrate the potential for continuously fitting data -- as it is collected -- in order to provide guidance on when enough information has been collected from the data to conclude the experiment. 



\begin{figure*}[t!] 
  \includegraphics[clip, trim=0 13.7cm 0 0, width=\textwidth]{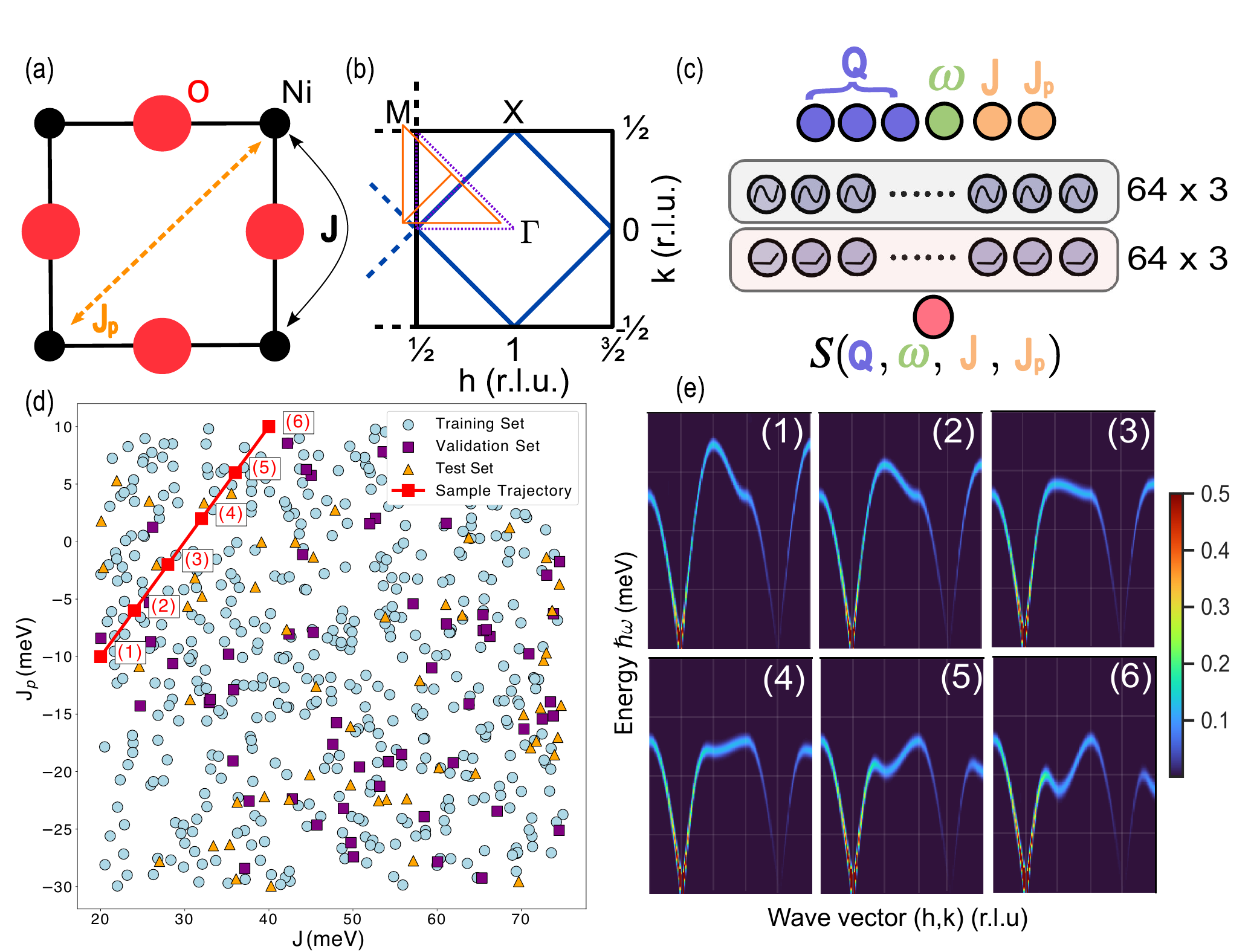}
  \caption{Overview of machine learning pipeline, model Hamiltonian and reciprocal space paths. (a) Ni$_4$O$_4$ square-lattice plaquette in La$_2$NiO$_4$. $J$ and $J_p$ are the first- and second-nearest-neighbor interactions.
  (b) The Brillouin zone for
the spin-1 square lattice magnetic structure. Selected high-symmetry points are
indicated. The two momentum paths are denoted by the purple and orange lines, respectively. (c) Visualization of the SIREN neural network for predicting the scalar dynamical structure factor intensity. All nodes exhibit a fully-connected architecture. The notation $64\times3$ and $64\times1$, represent three and one neural network layers with 64 neurons each. (d) Visualization of the distribution of training, test, and validation data in $J$-$J_p$ space. (e) Synthetic $S(\mathbf{Q},\omega)$ predictions from the SIREN model along the corresponding trajectory drawn in (d). Grid lines correspond to [0, 50, 100, 150, 200] and [$P$, $M$, $X$, $P$, $\Gamma$, $X$] for the energy and wave vector respectively.}
  \label{fig:fig_1}
\end{figure*}

\section{Machine Learning Approach}\label{section:MLapproach}
Our machine learning framework is based on implicit neural representations \cite{sitzmann2020implicit,xie2022neural}. These models are often described as coordinate networks as they take a coordinate as input and typically output a single scalar or a small set of scalars. For example, in the field of computational imaging, these networks learn mappings from pixel position ($i, j$) to RGB value. This representation allows for the model to be queried in between pixel values and implicitly represents the image through the trained weights of a neural network. We use a SIREN coordinate network~\cite{sitzmann2020implicit}, a fully-connected neural network~\cite{hastie2009elements} with sinusoidal activation functions, which has been shown to be able to accurately capture high-frequency features in images and scenes and has been particularly successful at tasks such as 3D-shape representation. Furthermore, gradients and higher-order derivatives of the mapping can be readily calculated and used for solving inverse-problems~\cite{sitzmann2020implicit,cheong2022novel,vlavsic2022implicit,levy2022amortized}. 

Here, the SIREN model acts as a fast and differentiable implicit representation for the hyper-volume of the dynamical structure factor across different model Hamiltonian parameters. The Hamiltonian we investigate in this work corresponds to the further nearest-neighbor Heisenberg model and is generally given by Eq. \ref{eqn:Hamiltonian}~\cite{coldea2001lco,marshall1971totns}.
\begin{align}
\label{eqn:Hamiltonian}
\mathcal{H}\!{}=\!{}J\sum_{\langle i,j\rangle}\!{}\hat{\mathbf{S}}_i\cdot\hat{\mathbf{S}}_j\!{}+\!{}J_p\sum_{\langle i,j'\rangle}\!{}\hat{\mathbf{S}}_i\cdot\hat{\mathbf{S}}_{j'}\!{},
\end{align}
As depicted in Fig.~\ref{fig:fig_1}a, $J$ and $J_p$ are the first- and second-nearest-neighbor
Heisenberg interactions on a square lattice. Thus for $Q_x$ and $Q_y$, a square lattice notation is utilized with $\mathbf{a}$ and $\mathbf{b}$ corresponding to the vectors connecting the first nearest neighbors.

Specifically, our SIREN model is trained to represent the scalar function $\mathrm{log} (1 + S(\mathbf{Q}, \omega, J, J_p))\in\mathbb{R}^1_+$, which is a logarithmic transformation of the dynamical structure factor evaluated at a specific $\mathbf{Q}\in\mathbb{R}^3$ (reciprocal lattice vectors in units of r.l.u.), $\hslash\omega\in\mathbb{R}^1$ (energy transfer in units of meV) and $J$ and $J_p\in\mathbb{R}^1$ (specific Hamiltonian coupling parameters). This functional mapping is described by Eq.~\ref{eqn:NN} and the SIREN model $\Phi$ is visualized in Fig.~\ref{fig:fig_1}c. 

\begin{equation}
\label{eqn:NN}
    \mathlarger{\Phi}(\mathbf{Q}, \omega, J, J_p) \rightarrow \mathrm{log}(1 + S(\mathbf{Q}, \omega, J, J_p))  
\end{equation}

We use a $\mathrm{log}(1 + x)$ transform in order to amplify a weak signal and to prevent ill-conditioned behaviour around zero. Here, we note that in principle, our model is written for three-dimensional $\mathbf{Q}$, however the neutron profiles in the subsequent sections do not include a $Q_z$ component due to limited scattering. The model is trained on 1,200 LSWT simulations of $S(\mathbf{Q}_{\mathrm{list}},\omega_{\mathrm{list}})$ over a large set of possible $J$, $J_p$ and on two paths in reciprocal space (Fig.~\ref{fig:fig_1}b). Here $\mathbf{Q}$-path 1 and 2 are represented as $P\rightarrow M\rightarrow X\rightarrow P\rightarrow \Gamma\rightarrow X$ and $P1 \rightarrow M 1 \rightarrow X1 \rightarrow P1 \rightarrow \Gamma 1 \rightarrow X1$ which corresponds to $\mathbf{Q}_{\mathrm{path1}} = \left\{\left[\tfrac{3}{4}\ \tfrac{1}{4}\ 0\right],\left[\tfrac{1}{2}\ \tfrac{1}{2}\ 0\right],\left[\tfrac{1}{2}\ 0\ 0\right],\left[\tfrac{3}{4}\ \tfrac{1}{4}\ 0\right],\left[1\ 0\ 0\right],\left[\tfrac{1}{2}\ 0\ 0\right]\right\}$ and $\mathbf{Q}_{\mathrm{path2}} =  \left[-0.07\ 0.03\ 0\right]+\mathbf{Q}_{\mathrm{path1}}$. Here, $\mathbf{Q}_{\mathrm{list}}\in\mathbb{R}^{N_\mathbf{Q}}$ and 
$\mathbb{R}^{N_\omega}$ is overloaded notation which refers to a series of $N_\mathbf{Q}$ and $N_\omega$ points in $(\mathbf{Q},\omega)$-space, respectively. 

Once the differentiable neural implicit model is trained, it is possible to use gradient-based optimization to solve the inverse problem of determining the unknown $J$ and $J_p$ parameters from data. Our objective function for the optimization task measures the Pearson correlation coefficient ($r$) of the logarithm of the predicted and the `true' $S(\mathbf{Q}, \omega, J, J_p)$ values (Eq.~\ref{eqn:Loss}). We use the logarithm to increase the weighting of weaker features in the data and we use the correlation as the metric because the normalization factors between the experiment and simulation are unknown. Using the prior is favourable as it enhances the weighting of the coherent excitation at high $\hslash\omega$ and further helps evade contamination due to statistical noise in the elastic and incoherent-inelastic scattering, primarily arising at low $\hslash\omega$ and which cannot be removed by background subtraction. The latter is important as we are not aiming to fully describe the spectral weights as this would require the exact handling of all individual neutrons in the full three-dimensional $\mathbf{Q}$-space instead of the averaged weight in the reduced two-dimensional $\mathbf{Q}$-space. During optimization, any subset of $(\mathbf{Q}_{\mathrm{list}},\ \omega_{\mathrm{list}})$ coordinates can be chosen as long as they fall along either of the paths defined in Fig.~\ref{fig:fig_1}b. Here, we note that from an inference point of view, any momentum or energy coordinates could be chosen, however our training data only includes two reciprocal space paths. To determine the Hamiltonian parameters, $J$ and $J_p$ are treated as free parameters in the optimization problem. The objective in Eq.~\ref{eqn:Loss} is optimized using the Adam optimizer~\cite{kingma2014adam}, a commonly used gradient-based optimization algorithm, exploiting the automatic differentiation capabilities in Tensorflow~\cite{abadi2016tensorflow} to calculate $\tfrac{\mathrm{d}L}{\mathrm{d}J}$ and $\tfrac{\mathrm{d}L}{\mathrm{d}J_p}$. See methods for further details.

\begin{equation}
\mathrm{L} = 1 - r(\mathrm{log}(1 + S_{\mathrm{measured}}), \mathlarger{\Phi}(\mathbf{Q}, \omega, J, J_p)) 
\label{eqn:Loss}
\end{equation}

In our method, it is not necessary to use all sets of $\mathbf{Q}_{\mathrm{list}}$, $\omega_{\mathrm{list}}$ along both paths to perform the fitting. Instead, random batches of coordinates $(\mathbf{Q}_{\mathrm{batch}}$, $\omega_{\mathrm{batch}})$ can be queried at each optimization iteration in order to improve computational efficiency and converge to a better minima, in a manner similar to the regularization effects of stochastic gradient descent~\cite{bottou2012stochastic}. Pseudo-code for the optimization procedure is provided in Algorithm \ref{alg:gradientdescent}.
\begin{algorithm}[H]
\begin{algorithmic}
\While{$\mathrm{N} < \mathrm{MaxIter}$}
    \State $\mathbf{Q}_{\mathrm{batch}},\ \omega_{\mathrm{batch}},\ S_{\mathrm{batch}}\ \sim\ [\mathbf{Q}_{\mathrm{list}},\ \omega_{\mathrm{list}},\ S_{\mathrm{list}}]$
    \State $\mathrm{log}(1 + S_{\mathrm{pred}}) = \mathlarger{\mathrm{\Phi}}(\mathbf{Q}_{\mathrm{batch}},\ \omega_{\mathrm{batch}},\ J,\ J_p)$
    \State $J,\ J_p \gets \mathrm{ADAM}(\mathrm{L}(S_{\mathrm{batch}},\ S_{\mathrm{pred}}))$
\EndWhile
\end{algorithmic}
\caption{Differentiable Surrogate Optimization}
\label{alg:gradientdescent}
\end{algorithm}

Although our approach uses a neural network forward model and differentiable optimization, we note that in many machine learning studies, the default approach is to train an inverse model on simulated data. More specifically a model which takes raw data and directly predicts the unknown parameters. In our case, this would be a model of the form $S(\mathbf{Q}_{\mathrm{list}}, \omega_{\mathrm{list}})\rightarrow (J,\ J_p)$. Here, for example, $S(\mathbf{Q}_{\mathrm{list}}, \omega_{\mathrm{list}})$ could be represented as an image and the prediction task could use standard convolutional neural network architectures~\cite{aloysius2017review}. This is similar to the work in Refs.~\cite{butler2021interpretable,anker2022using} which train inverse models to predict the spin model class from experimental data. While such approaches can often work well on simulated test data, inference is often much more challenging for experimental data and often requires detailed modelling and corrective dataset augmentations of experimental effects accounting for attributes such as background noise, missing data, and matching instrumental profiles~\cite{chitturi2021automated, wang2020rapid}. Specifically, from an inverse modeling standpoint, the experimental data is generally outside the distribution used to train the model. Here, generative approaches, such as the recently proposed cycle-GAN model in Ref.~\cite{anker2022using}, offer an elegant alternative to minimize the deviation between experimental and simulated data. In cases where the predicted outcomes have continuous values (i.e. regression tasks), inverse prediction pipelines can be even less robust to data distribution differences as the model may have to learn more subtle features. For this reason, instead of choosing an inverse modelling approach, we opted for a SIREN neural forward model with differentiable optimization. For this approach, the machine learning model only makes predictions in-distribution and the experimental non-idealities are considered only in the optimization step. Here, we note that the generative approach in Ref.~\cite{anker2022using} could likely augment our method by using the "translated" experimental data directly in the loss function. 

\section{Results and Discussion}

\begin{figure}[!h]
\centering 
\includegraphics[clip, trim=0 2.5cm 0 0, width=\linewidth]{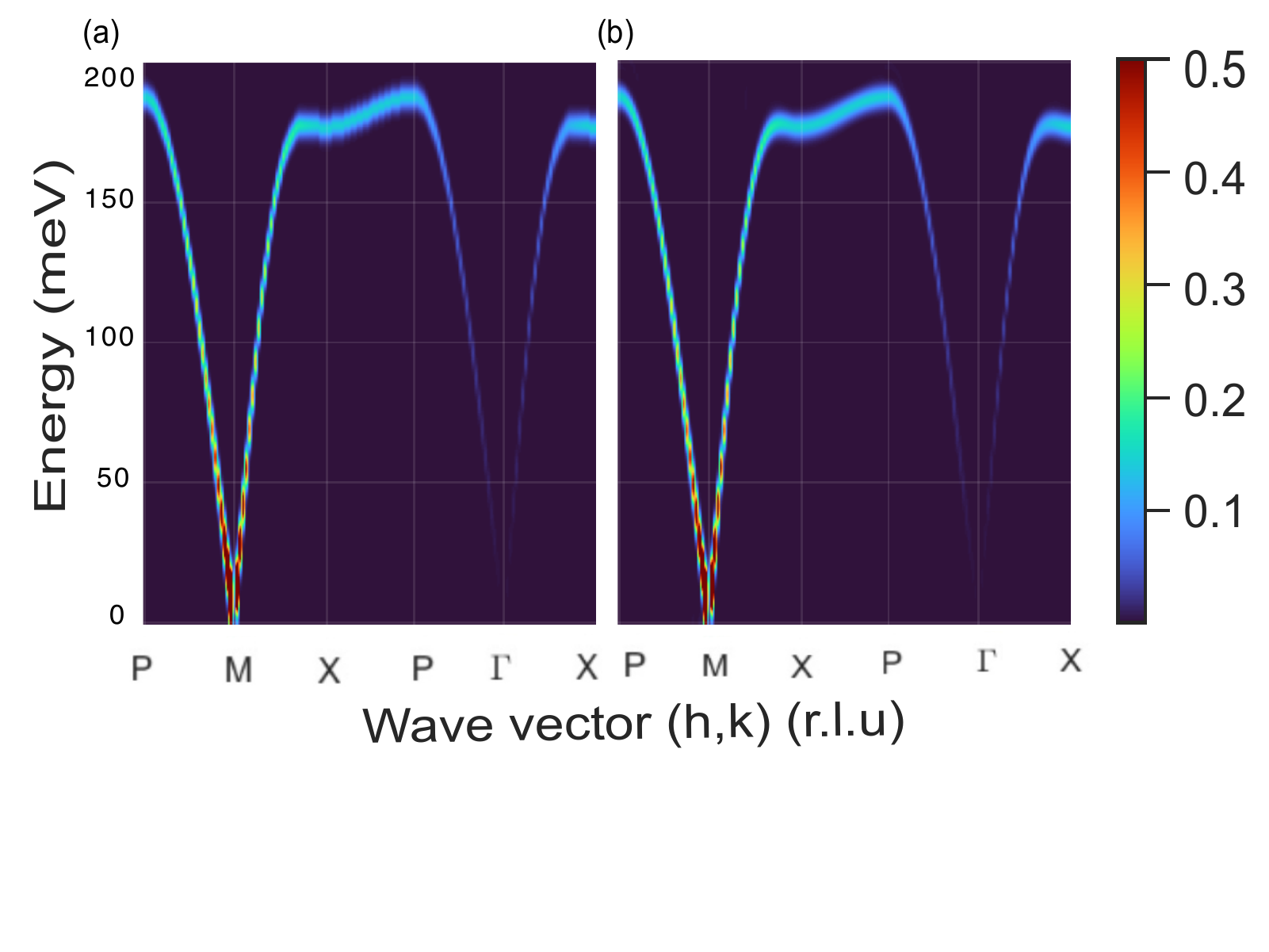}
\caption{Visualization of ground-truth linear spin wave theory (LSWT) simulation and corresponding machine learning forward model. Example of ground-truth simulated S(\textbf{Q}, $\omega$) calculated using the SpinW software program (a) and corresponding machine learning forward model prediction (b) given parameters $J$ = 45.57, $J_p$ = 2.45 meV.}
\label{fig:simulated_pred}
\end{figure}

We first characterized the performance of the machine learning framework on simulated SpinW data in order to demonstrate the viability of using a neural implicit model as a surrogate for the LSWT simulator. Fig.~\ref{fig:simulated_pred} shows the LSWT simulation and machine learning "simulation" with input parameters of $J= 45.57$\,meV and $J_p =  2.45$\,meV. Note that in this example, the machine learning framework was fed ($J,\ J_p$) directly (instead of obtaining these parameters using gradient descent through the neural representation). Evidently, the machine learning prediction and LSWT simulation are almost indistinguishable, highlighting the ability of the neural representation to mimic the theoretical calculation. 

\begin{figure*}[tb] 
    \includegraphics[clip, trim=0 16cm 0 0, width=\textwidth]{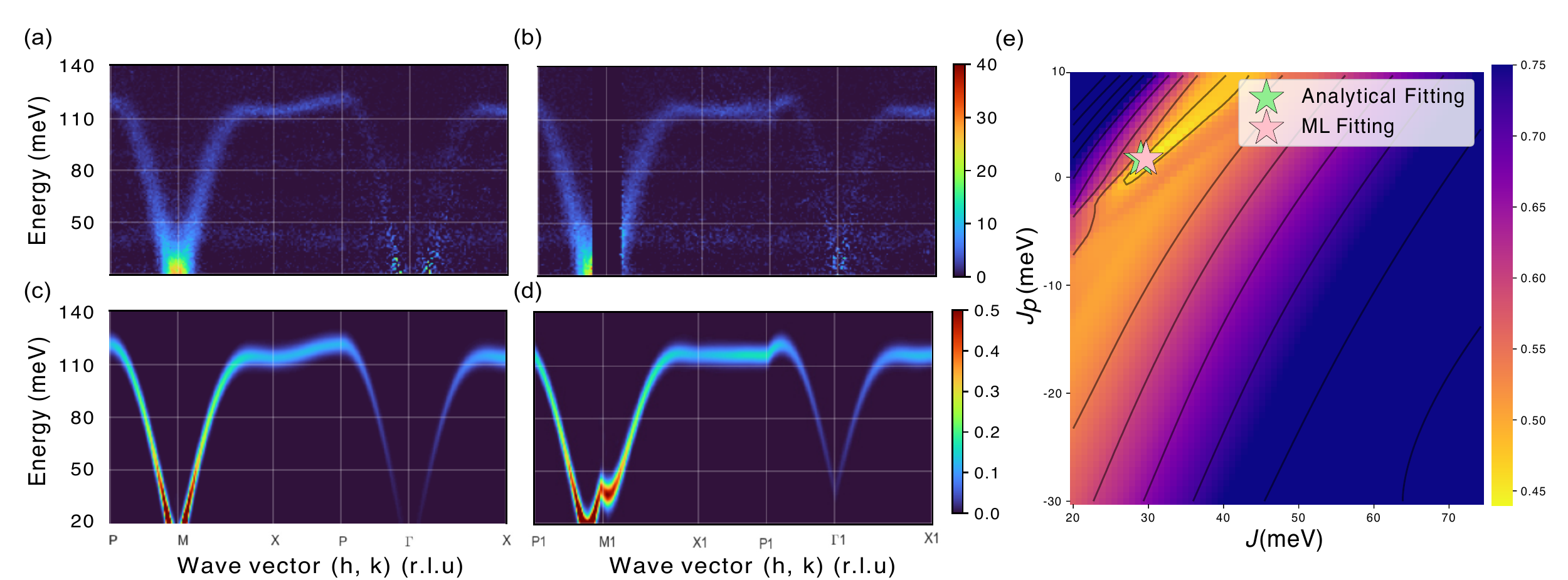}
  \caption{Machine learning forward model prediction and gradient-based optimization. (a) and (b) show experimental data after automated background subtraction. The color bars reflect $S(\mathbf{Q}, \omega)$ in units of: $ \mathrm{mbarn}$ $\mathrm{sr}^{-1} \mathrm{meV}^{-1} \mathrm{f.u.}^{-1}$. (c) and (d) show corresponding machine learning predictions for both paths. The visual predicted profiles are in close agreement with the experimental data. Deviations at low $\hslash\omega$ are due to the neglection of the anisotropy spin gaps in our model. (e) Visualization of the loss landscape for objective fitting in Hamiltonian parameter space ($J$, $J_p$).}
  \label{fig:fig_2}
\end{figure*}

Although our model can clearly approximate simulated data well, the main motivation of this approach is to provide a tool that can reliably extract the spin Hamiltonian parameters of interest from real, experimental data. For this reason, we applied our method to measured inelastic neutron scattering data, after automatic background-subtraction, on a quasi-2D N\'eel antiferromagnet La$_2$NiO$_4$ collected at the Spallation Neutron Source~\cite{granroth2010sequoia}. Though a full 3D dataset was collected, we chose two paths in $\mathbf{Q}$-space to simulate spectra for the model training prior to any inclusion of the real data. After the model had been trained on both simulated paths, we used gradient-based optimization to solve the inverse problem of determining $J$ and $J_p$ from the data. Here, we note that the optimization for both experimental paths was performed jointly and therefore the fit parameters are the same for both. We found that our approach yields excellent predictions, both qualitatively and quantitatively, relative to the results of a detailed and expensive analytical fit, as shown in Fig.~\ref{fig:fig_2}a and b. The analytical parameters in the LSTW limit adapted from Petsch \textit{et al.}~\cite{petsch2023highenergy} are $J = 29.00(8)$\,meV and $J_p=1.67(5)$\,meV. In addition, the experimental data show spin gaps at low energy which are here neglected. The parameters obtained from the ML fitting are $J = 29.68$\,meV and $J_p= 1.70$\,meV. The small overestimation of $J$ arises due to a number of factors. Firstly, there exist small differences due to the 3-dimensionality of $\mathbf{Q}$ and hence, in the following variations in the magnetic form factors and polarization factors. The 3-dimensional information is not included in the analyzed data, as they are averaged over $Q_z\in[-10,10]$\,r.l.u., and thus these factors are neglected in the simulated images. In addition, the resolution function and finite lifetime are only approximations here and further, any multi-magnon scattering is not described by LSWT. Finally, the experimentally observed energy shift by the spin gaps~\cite{petsch2023highenergy,Nakajima1993la2nio4} is not considered to minimize the number of parameters for clarity. Note, we also experimented with fitting each path independently and also obtained similar predictions; for path two, this is a notable achievement since a significant section of the data is missing in the experiment (See SI2). In addition, in SI3, we provide fitting results from SpinW with algorithmic peak-fitting, which yields similar results for this dataset. 

In addition, since the neural implicit model is cheap to evaluate, we also constructed a loss landscape of the objective function with respect to $J$ and $J_p$. We see that the objective function is well-behaved and that the gradient descent scheme finds a fit close to the analytical result (Fig.~\ref{fig:fig_2}e).

Here, we emphasize that the only information provided to the algorithm is knowledge of a region of $(\mathbf{Q},\hslash\omega)$-space from which to perform automatic background subtraction prior to fitting the data. Importantly, no peak finding or extraction is needed as the optimization objective uses the intensity of all provided voxels in the $(\mathbf{Q},\hslash\omega)$-space or pixels on the 2D intensity map, respectively, rather than magnon peak positions $\hslash\omega_\mathbf{Q}$. We expect that the ability to fit such complex data in real-time could be readily coupled to autonomous experiment steering agents. For example, the neural implicit model can provide fast and scalable forward computations to provide sufficient sampling for accurate distribution estimations, which are essential in Bayesian experiment designs~\cite{granade2012robust, mcmichael2022simplified}.

\begin{figure*}[t!] 
\includegraphics[clip, width=\textwidth]{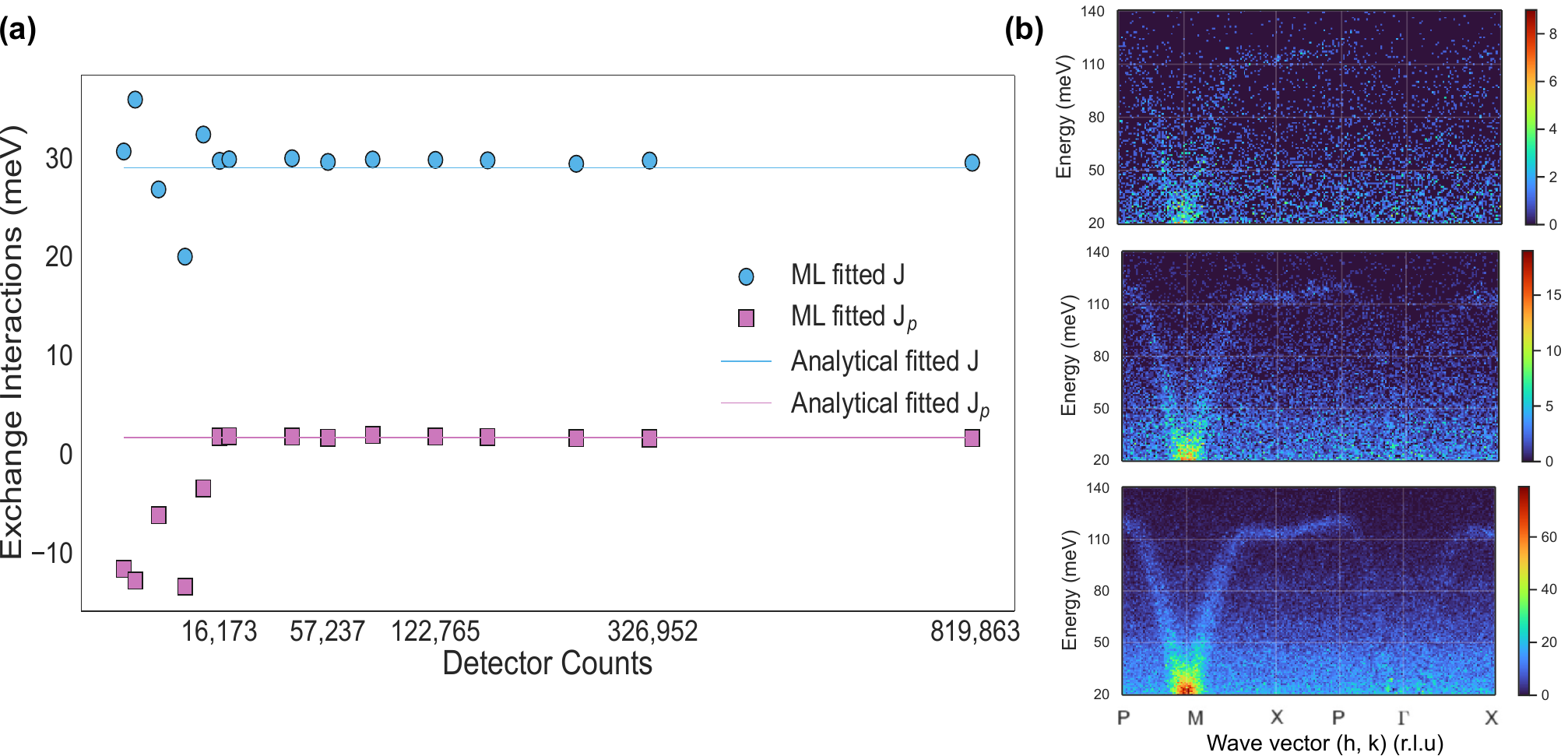}
  \caption{(a) Machine learning prediction for $J$ and $J_p$ as a function of detector count with square-root scaling of the `Detector Counts' axis in regard to Poisson statistics. Note, the machine learning prediction converges much earlier than the count-time recorded in the experiment. (b) Visualization of plausible low-count (without algorithmic background subtraction) data with total detector counts: 16,173, 57,237, and 326,952 (top to bottom).}
  \label{fig:fig_3}
\end{figure*}

In real experimental settings, another critical aspect is the ability to make rapid decisions on whether sufficient statistics have been obtained for understanding the necessary physics being measured. Especially since neutron scattering measurements typically have low detector count rates, this is a major influence on the efficiency of measurement time at facilities. Moreover, one would like to minimize the amount of time needed without sacrificing data quality, or rendering information statistically insignificant, and ultimately reliably analyze magnetic excitations in the most efficient way. 

To probe the effectiveness of our framework for real-time fitting during an experiment, and to reduce data collection time, we used the current experimental data to generate plausible data for low counting situations. Specifically, we smooth the experimental data and use it as a probability distribution which is sampled using rejection sampling (See Methods). Here, we note that there is no “detector noise” and that the noise in the experiment comes purely from the background scattering of the data.

In Fig.~\ref{fig:fig_3}a, we show the obtained parameters from the machine learning fitting as a function of the number of detected neutrons within the path region. Visualizations of path 1 at selected points in time are also shown in \ref{fig:fig_3}b. The machine learning prediction is obtained as the lowest objective value from 10 independent gradient descent optimizations starting from random locations in Hamiltonian parameter space. We further note that using the median prediction also gives very similar results. This test demonstrates that the machine learning model quickly converges to the true solution.

The ability to continuously fit data as it is collected is very useful from an experimental point of view. Had other paths in reciprocal space been available, it would have simply required training with additional simulations and without any changes to the overall machine learning model or framework. This is because the model was built to generically accept reciprocal coordinates as input. In general, fitting this type of high-dimensional data is not compatible with having to run peak finding algorithms (which are often manually guided) for each path. Importantly, the algorithm described here will be a valuable tool for both carrying out these types of experiments faster, or enabling multiple experiments to be performed. From this analysis, we demonstrate the ability of the machine learning model to be trained prior to an experiment to allow for real-time fitting and decision making. 

\section{\label{sec:conclusions}Conclusions}

We developed a powerful tool for identifying the key parameters that describe the linear spin wave spectrum. Besides LSWT, our machine learning algorithm can be combined with more complex and time-consuming models derived from e.g. exact diagonalization, and density-matrix renormalization group methods. It breaks the barrier of real-time fitting of inelastic neutron and x-ray scattering data, bypassing the need for complex peak fitting algorithms or user-intensive post-processing, and allows for the incorporation of entirely three-dimensional data in reciprocal space. Furthermore, the ability to fit data continuously throughout an experiment will be useful for determining an optimal stopping point for data collection as well as for guiding experiments. The approach opens up new 
opportunities which can significantly improve the ability to analyze scattering from excitations in ordered quantum systems.

\section*{Methods}
\label{section:Methods}

\subsection*{Sample Preparation and Data Collection}
\label{section:MethodsExperiment}
In the experiment a 21\,g single crystal of the quasi-2D N\'eel antiferromagnet La$_2$NiO$_{4+\delta}$ ($P4_2/ncm$ with $a=b=$5.50\,\AA, $c=$12.55\,\AA), grown by the floating-zone technique, was utilized. The presented time-of-flight neutron spectroscopy data were collected on the SEQUOIA instrument at the Spallation Neutron Source at the Oak Ridge National Laboratory~\cite{granroth2010sequoia} with an incident neutron energy of 190\,meV, the high-flux Fermi chopper spun at 300\,Hz and a sample temperature of 6\,K. The data is integrated over the out-of-plane momentum $Q_z\in\pm10$\,r.l.u. The lattice can be approximated by $I4/mmm$ with $a=b\approx3.89$\,\AA. $Q_x$ and $Q_y$ for $I4/mmm$ are equivalent to $Q_x$ and $Q_y$ in the square-lattice notation. For more details see Ref.~\cite{petsch2023highenergy}.


\subsection*{SpinW Simulation and Fitting}
\label{section:MethodsSpinW}
The two momentum paths used for $S(\mathbf{Q},\omega)$ simulation are $\mathbf{Q}_{\mathrm{list1}} = \left\{\left[\tfrac{3}{4}\ \tfrac{1}{4}\ 0\right],\left[\tfrac{1}{2}\ \tfrac{1}{2}\ 0\right],\left[\tfrac{1}{2}\ 0\ 0\right],\left[\tfrac{3}{4}\ \tfrac{1}{4}\ 0\right],\left[1\ 0\ 0\right],\left[\tfrac{1}{2}\ 0\ 0\right]\right\}$ and $\mathbf{Q}_{\mathrm{list2}} =  \left[-0.07\ 0.03\ 0\right]+\mathbf{Q}_{\mathrm{list1}}$, respectively. When plotting spin wave spectra using the SpinW software~\cite{toth2015linear}. 600 simulations were performed for each path (1200 total) corresponding to randomly sampling $J$ and $J_p$ in ranges [20, 75] and [-30, 10] meV. The lower limit for $J$ and upper limit for $J_p$ are chosen such that the ground state remains the N\'eel state which is satisfied in LSWT for $J>2J_p$ and $J>0$. For each location in $\mathbf{Q}$, the corresponding energies from 0 - 200\,meV were obtained. The quantum fluctuation renomalization factor $Z_c$ is set to 1.09~\cite{igarashi19921, singh1989thermodynamic,petsch2023highenergy}. After simulation, the data was convoluted with an energy-dependent kernel based on the beamline instrument profile. For this procedure, an in-built tool from SEQUOIA was used to give a polynomial fit for the dependence of the resolution (FWHM) in meV on the energy transfer ($\hslash\omega$) in meV: $\mathrm{FWHM} = 1.4858\times10^{-7}(\hslash\omega)^3 + 1.2873\times10^{-4} (\hslash\omega)^2 - 0.084492 \hslash\omega +14.324$~\cite{granroth2010sequoia}. In addition, the data was broadened with a 1D Gaussian kernel $(\sigma = 5$ pixels) in $\mathbf{Q}$ to correct for the discrete sampling of the simulation and to partially consider the momentum resolution of the instrument.

The SpinW-software-based spin wave spectrum fitting was implemented using its built-in function. The inputs are peak information extracted from experimental spin-wave dispersion data. The $R$ value is optimized using a particle swarm algorithm to find the global minimum defined as $R=\sqrt{1/n_E\times \sum_{i,\mathbf{Q}}1/\sigma_{i,q}^2(\hslash\omega_{i,\mathbf{Q}}^{\mathrm{sim}}-\hslash\omega_{i,q}^{\mathrm{meas}})^2}$, where $(i,q)$ index the spin-wave mode and momentum, respectively. $E_{sim}$ and $E_{meas}$ are the simulated and measured spin wave energies, $\sigma$ is the standard deviation of the measured spin-wave energy determined previously by fitting the inelastic peak and $n_E$ is the number of energies to fit.

\subsection*{Surrogate Model Training}
\label{section:MethodsNN}

A 5-layer SIREN neural network (Fig.~\ref{fig:fig_1}c) was trained on 1,000 simulations of ($S(\textbf{Q}, \omega)$, $J$, $J_p$) tuples; 200 simulations were left aside for validation and testing. Here, $\hslash\omega$ [0 - 200]\,meV, $J$ [20 - 75]\,meV and $J_p$ [-30 - 10]\,meV were normalized to 0-1 in order for all the parameters to be on approximately the same scale. The model was trained to predict $\text{log} (1 + S(\mathbf{Q}, \omega, J, J_p))$ by optimizing the mean-squared-error objective $L$ between the prediction and the label with respect to the network parameters. During training, the following hyper-parameters and settings were used: Adaptive Moment Estimation (ADAM) algorithm for optimization ($\beta_1 = 0.9$, $\beta_2 = 0.999$)  \cite{kingma2014adam}, batch size = 2,048, learning rate = 0.001. The learning rate was exponentially decayed by a factor of $\text{exp}(-0.1)$ for every epoch after the first ten epochs. We used NVIDIA A100 GPU hardware with the Keras API \cite{chollet2015keras} and the model was trained for 50 epochs. 

\subsection{\label{sec:MLPE}{Machine Learning Parameter Extraction}}
\label{section:MethodsOptimize}

Prior to differentiable optimization, the experimental data were automatically background subtracted using the following procedure. First, a region of $(\mathbf{Q}_{\mathrm{list}},\ \omega_{\mathrm{list}})$ space was chosen for each slice ({160-170 pixel location in the Q-axis}) and averaged across $\mathbf{Q}_{\mathrm{list}}$ to yield a one-dimensional energy profiles. This procedure was chosen based on prior assumptions on the isotropic nature of the scattering and the N\'eel ground state. Next, the one-dimensional energy profiles were fit using a Savitzy-Golay filter (window size = 51, polynomial order = 3) and used for background subtraction. 

The unknown $J$, $J_p$ parameters were recovered from data using gradient-based optimization of the neural network implicit representation. For the experimental data presented in this work, the metric $(1 - r)$ between the measured and simulated $(1 + S(\mathbf{Q},\ \omega,\ J,\ J_p))$ was used as the objective function (Eq. \ref{eqn:objective}; here, $r$ refers to the Pearson correlation coefficient. No normalization was performed for scaling the simulation data relative to the experimental data. 
\begin{equation}
\label{eqn:objective}
L = 1 - r(\mathrm{log}(1 + S_{measured}), \mathlarger{\Phi}(\mathbf{Q}, \omega, J, J_p)) 
\end{equation}

The objective $L$ was optimized using the ADAM algorithm with respect to $J$ and $J_p$ and $\mathbf{Q}_{\mathrm{list}}$ and $\omega_{\mathrm{list}}$ were randomly sampled from the list of paths containing the experimental data. Here, a batch size of 4,096 was used for the$(\mathbf{Q}_{\mathrm{list}},\ \omega_{\mathrm{list}})$ sampling, with 2,000 Adam optimization steps and a learning rate of 0.005. 

\subsection*{Low count data generation and fitting}
\label{section:MethodsLowData}

High-count data for each slice (without background subtraction) were smoothed using a 3x3 Gaussian convolutional kernel. The resultant images were each normalized to (0, 1) using the total intensity. Each slice was treated as a probability distribution which was sampled using Monte-Carlo rejection sampling. This process was used to create a series of datasets with neutron counts in the range $(1\!{}\times\!{}10^4$ - $9\!{}\times\!{}10^6)$. Each dataset was individually and automatically background subtracted by the previously described method and fit ten times from random starting locations in ($J$, $J_p$) using the machine learning optimization procedure. Note, the corresponding low-count data was used in order to perform the automated background subtraction. 

\section*{Data Availability}

The simulation data and code are available at~\url{https://github.com/slaclab/neural-representation-sqw.git}.

\section*{Author Contributions}

S.R. Chitturi, Z. Ji and A.N. Petsch contributed equally to this work.

\section*{Acknowledgements}

This work is supported by the U.S. Department of Energy, Office of Science, Basic Energy Sciences under Award No. DE-SC0022216, as well as under Contract DE-AC02-76SF00515 both for the Materials Sciences and Engineering Division and for the Linac Coherent Light Source (LCLS). A portion of this research used resources at the Spallation Neutron Source, a DOE Office of Science User Facility operated by the Oak Ridge National Laboratory. J. J. Turner acknowledges support from the U.S. DOE, Office of Science, Basic Energy Sciences through the Early Career Research Program. Z. Ji is supported by the Stanford Science fellowship, and the Urbenek-Chodorow postdoctoral fellowship awards.  A.N. Petsch and S.M. Hayden acknowledge funding and support from the Engineering and Physical Sciences Research Council (EPSRC) under Grant Nos. EP/L015544/1 and EP/R011141/1.


\nocite{*}

\bibliography{apssamp}

\clearpage
\onecolumngrid

\appendix

\section{SI1: Data without background subtraction}

\begin{figure}[H] 
    \includegraphics[clip, trim=0 16cm 0 0, width=\textwidth]{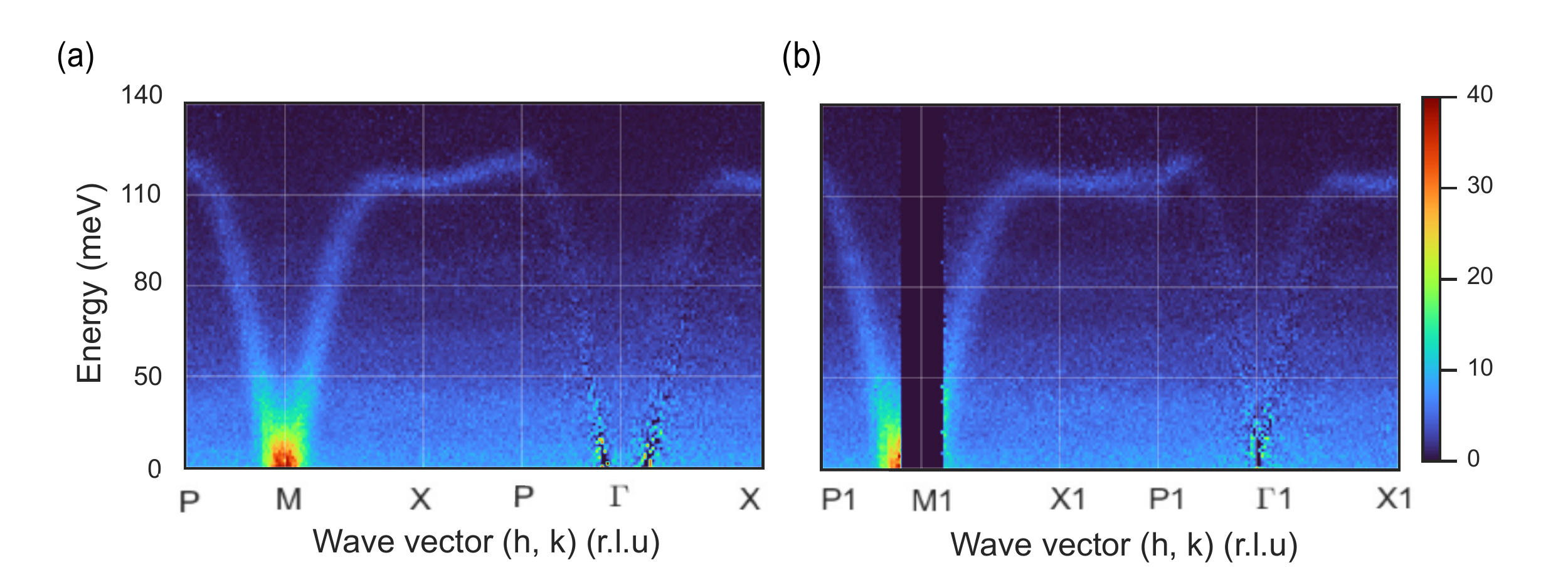}
  \caption{Inelastic neutron scattering dataset without automated background subtraction. Visualizations are shown for (a) path 1 and (b) path 2. Note, path 2 contains a portion of missing data.}
  \label{fig:fig_wbkg}
\end{figure}

Visualization of path 1 (a) and path 2 (b) for inelastic neutron scattering data without automated background subtraction (Fig.~\ref{fig:fig_wbkg}). Note, in path 2, a region of the data is missing.

\newpage
\section{SI2: Missing Region Interpolation}

In the raw data for path 2, a region is missing. However, since the machine learning model is a forward surrogate for the LSWT simulation, it is able to make predictions even where there is no data. Note, that this is a significant advantage over the inverse modelling approach.

For the displayed path there is no data available in the here utilized neutron dataset for the missing region of $\mathbf{Q}$-space. However, data is available for $(Q_x,Q_y)\mapsto(Q_y,Q_x)$, so for the 2D momentum rotated by 45$^\circ$ of the missing path region. As this compound is assumed to be fully twinned $S(Q_x,Q_y,\omega)=S(Q_y,Q_x,\omega)$ and thus, the missing region in $\mathbf{Q}$ can be substituted by the equivalent data with $Q_x$ and $Q_y$ exchanged. The data along the full path with the missing region substituted is depicted in comparison with the result predicted by our forward model. Note, in this case, the prediction for path 2 only uses the data from path 2. No information from path 1 is utilized in the fitting. 

\begin{figure}[H]
\centering 
  \includegraphics[clip, trim=0 23cm 0 0, width=\textwidth]{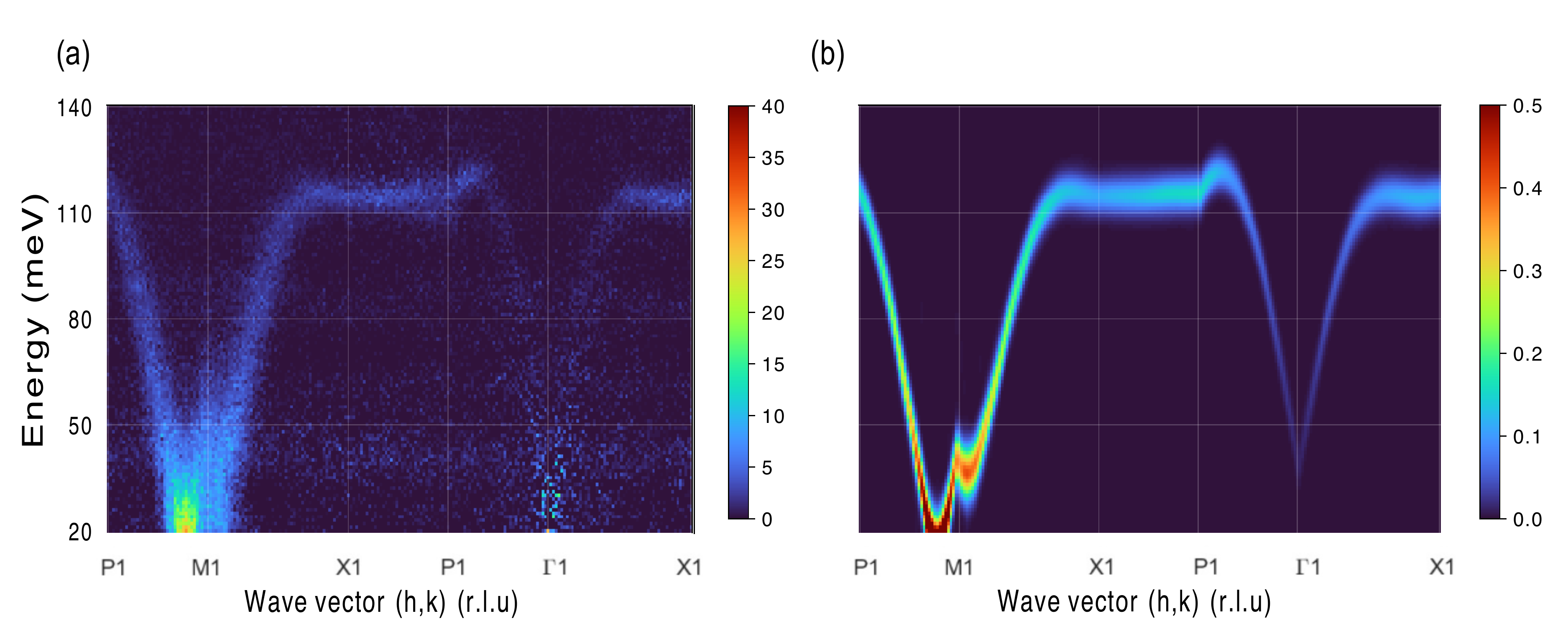}
\caption{Machine learning forward model accurately predicts scattering profile for missing region. (a) Experimental data with missing region filled in by exchanging $Q_x$ and $Q_y$. (b) ML prediction using only experimental data from path 2 with missing region. Evidently, the ML prediction closely models the true experimental data.}
\label{fig:sim}
\end{figure}

\newpage
\section{SI3: Results from SpinW fitting}
\begin{figure}[H]
\centering 
  \includegraphics[clip, trim=1.5cm 8.5cm 1.5cm 9cm, width=\textwidth]{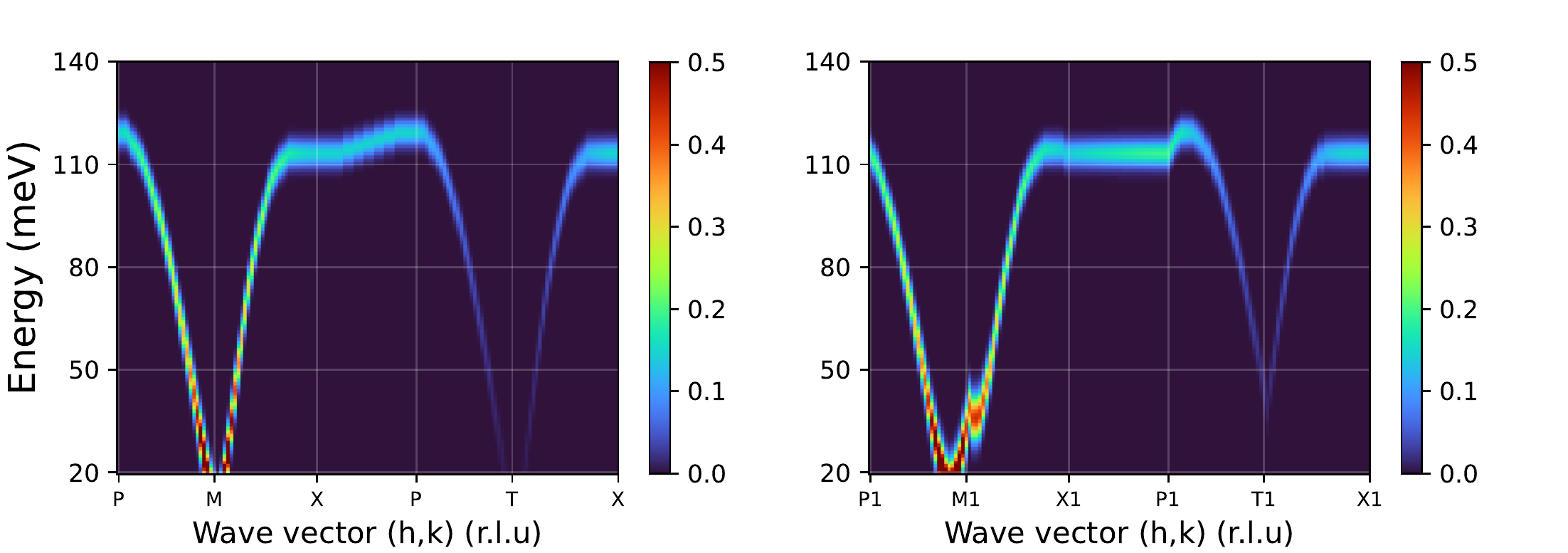}
\caption{SpinW fitting result for path 1 (left) and 2 (right) as a benchmark.}
\label{fig:spinw}
\end{figure}

This section shows the SpinW fitting results for the $(\rm{J},\rm{J_p})$ parameters when both path 1 and path 2 are used for the fitting (Fig. 3a and b).  SpinW fitting was carried out to provide a benchmark for the ML results: $(\rm{J},\rm{J_p})=(29.15,1.55)$ (The result is the median of five fits with 100 maximum iterations each, path 1 resulting result is shown in Fig.~\ref{fig:spinw}). Relative to the ML method, SpinW fitting does not utilize all the available pixel information and instead requires additional peak finding and peak fitting steps. When the spin Hamiltonian has more parameters, or when there are multiple eigenmodes in the energy region of interest, our ML algorithm may show greater advantage both in time and accuracy. 

\end{document}